\documentclass[reprint,amssymb,amsmath,aps,unsortedaddress,superscriptaddress, showpacs, footinbib,prb]{revtex4-1}
\usepackage{graphics}
\usepackage{graphicx}
\usepackage{amssymb}
\usepackage{color}

\usepackage{amsmath}

\begin{document}
\title{Antiferromagnetism of Ni$_2$NbBO$_6$ with \textit{S}~=~1 dimer quasi-1D armchair chains}

\author{G. Narsinga Rao}
\affiliation{Center for Condensed Matter Sciences, National Taiwan University, Taipei 10617, Taiwan}

\author{Viveka Nand Singh}
\affiliation{Institute of Atomic and Molecular Sciences, Academia Sinica, Taipei 10617, Taiwan}
\affiliation{Department of Physics, National Taiwan University, Taipei 10617, Taiwan}

\author{R. Sankar}
\affiliation{Center for Condensed Matter Sciences, National Taiwan University, Taipei 10617, Taiwan}

\author{I.~Panneer~Muthuselvam}
\affiliation{Center for Condensed Matter Sciences, National Taiwan University, Taipei 10617, Taiwan}

\author{Guang-Yu Guo}
\email{gyguo@phys.ntu.edu.tw}
\affiliation{Department of Physics, National Taiwan University, Taipei 10617, Taiwan}

\author{F. C. Chou}
\email{fcchou@ntu.edu.tw}
\affiliation{Center for Condensed Matter Sciences, National Taiwan University, Taipei 10617, Taiwan} \affiliation{National Synchrotron Radiation Research Center, Hsinchu 30076, Taiwan} \affiliation{Taiwan Consortium of Emergent Crystalline Materials, Ministry of Science and Technology, Taipei 10622, Taiwan}

\begin{abstract}

Long range antiferromagnetic (AFM) ordering of Ni spins in Ni$_{2}$NbBO$_{6}$ has been studied with single crystal from spin susceptibility measurement and compared with the \textit{ab initio} calculation results consistently. Below ${\it T}_\mathrm{N}$$\sim$23.5~K, the \textit{S}~=~1 spins align along the $a$-direction for edge-shared NiO$_6$ octahedra which form crystallographic armchair chains along the $b$-direction. The isothermal magnetization ${\it M}(H)$ below ${\it T}_\mathrm{N}$ shows spin-flop transition for magnetic field above $\sim$~36~kOe along the $a$-axis, which indicates the spin anisotropy is along the $a$-direction. The electronic and magnetic structures of Ni$_2$NbBO$_6$ have also been explored theoretically using density functional theory with generalized gradient approximation plus on-site Coulomb interaction (U). These calculations support the experimentally observed antiferromagnetism of Ni$_{2}$NbBO$_{6}$. In particular, the long range AFM ordering below ${\it T}_\mathrm{N}$ can be dissected into armchair chains which consists of \textit{S}~=~1 dimers of $J_{2}$$\sim$2.43~meV with ferromagnetic (FM) intra-chain and inter-chain couplings of size $\lesssim$$\frac{1}{2}$$|J_{2}|$.

\end{abstract}


\maketitle
\section{Introduction}
Low dimensional magnetic materials have attracted considerable attention due 
to their interesting low temperature properties with the involved strong quantum fluctuations~\cite{Diep, Troyer}. 
Extensive studies of materials with geometric frustration on square, 
triangular, zigzag chains and zigzag ladders spin systems have been explored for the 
diverse magnetic ground states. The zigzag spin chain of \textit{S}~=~1/2 with 
antiferromagnetic (AFM) interactions between nearest neighbor (NN) and next nearest 
neighbor (NNN) is about the most commonly studied frustrated system~\cite{White, Matsuda}. In zigzag spin 
chain system with \textit{S}~=~1, the ground state phase diagram as a function of 
anisotropy and ratio between NN and NNN interactions exhibits different 
phases~\cite{Hikihara, Kolezhuk}. Metal borates are expected to be good candidates 
to serve as links for transition metal polyhedra giving rise to different 
low-dimensional structures~\cite{Grice}. Another important role of the borate anions, being non-magnetic, is to allow transmission of magnetic interactions via a super-superexchange route~\cite{Grishachev, Attfield, Fernandes, Fernandes1}. 
 
	In the present work, we report the crystal growth and the magnetization measurement results along the three principal 
directions of Ni$_{2}$NbBO$_{6}$. Crystallographically Ni$_{2}$NbBO$_{6}$ has been found to be a  \textit{S}~=~1 armchair spin chain 
system~\cite{Ansell}. We found that a long range AFM spin ordering exists below ${\it T}_\mathrm{N}$$\sim$23.5~K. A sizable inter-chain coupling leads to the 3D long range AFM spin ordering with an on-site anisotropy along the $a$-direction, which is as confirmed by the field-induced spin flop transition. We also studied the electronic and 
magnetic properties of Ni$_2$NbBO$_6$ within the density functional theory 
with the generalized gradient approximation. We found that the system 
consists of unconventional armchair chains which are formed with ferromagnetically coupled \textit{S}~=~1 dimers with intra- and inter-chain coupling constants which are nearly half of that for the \textit{S}~=~1 dimer. An interpretation on the experimental observation on the AFM and spin flop transition is provided and compared with the calculated results.

\section{Experimental and computational details}

\begin{figure}[t]
\centering
\includegraphics[width=0.4\textwidth]{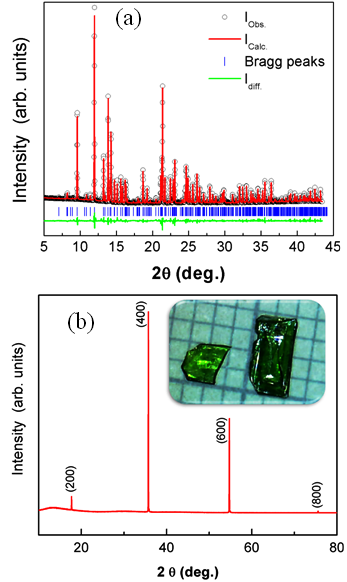}
\caption{(Color online) (a) The powder XRD pattern  
from the pulverized as grown Ni$_{2}$NbBO$_{6}$ crystal. (b) The XRD pattern of 
 Ni$_{2}$NbBO$_{6}$ crystal perpendicular to the large surface to show (h00) peaks with preferred orientation. Inset is the as grown single crystal.}
\label{fig:Figure1}
\end{figure}

Single crystal of Ni$_{2}$NbBO$_{6}$ was grown by a
flux method using borate as the solvent. A mixture of 6.6~g NiO, 20~g 
Nb$_{2}$O$_{5}$ and 33~g of Na$_{2}$B$_{4}$O$_{7}$ 
were placed in a platinum crucible and heated to 
$1250^{\circ}$C in a box furnace for 24~hours. The 
furnace was slowly cooled down to $850^{\circ}$C at 
a rate of $3^{\circ}$C/h and then cooled down to 
room temperature at the rate of $80^{\circ}$C/h. 
The single crystals in green color (shown in the inset of 
Fig.~\ref{fig:Figure1}(b)) were separated from the 
borate flux by leaching with a dilute solution of 
HNO$_{3}$. The crystal structure and phase purity of the samples were checked by powder X-ray diffraction (XRD) using the synchrotron X-ray of $\lambda$~=~0.619~\AA (NSRRC, Taiwan) at room temperature. The field cooled (FC) and 
zero field cooled (ZFC) magnetization curves were 
measured in a commercial Vibrating Sample 
Magnetometer (VSM, Quantum Design, USA) from 1.8~K 
to 300~K in the presence of various applied 
magnetic fields. The isothermal magnetization 
({\it M}) data were also recorded at selected 
temperatures.

Theoretical calculations have been performed 
based on first-principle density
functional theory (DFT) with generalized gradient
approximation (GGA)~\cite{gga}. The on-site Coulomb
energy U has been taken into account using the
GGA+U scheme~\cite{gga+U}. We have used effective
$U_{eff}=(U-J)=6$~eV for the Ni atoms in the GGA+U calculations.
We used the accurate full-potential projector-augmented wave (PAW) 
method~\cite{vasp_1} implemented in the Vienna 
{\it ab initio} simulation package 
(VASP)~\cite{vasp_2,vasp_3, vasp_4}. 
Experimental lattice parameters were used in the calculation. The primitive unit cell contains four 
Ni$_2$NbBO$_6$ formula units. 
In the present calculations, we used the tetrahedron method with 
Bl\"{o}chl corrections for the Brillouin zone integration with a 
$\Gamma$-centered Monkhorst-Pack k-point mesh of $(8\times10\times18)$.
A large plane wave cutoff energy of 500~eV was taken, and the 
convergence criterion for the total energy was $10^{-6}$~eV. 

%---------------------------------------------------------------
%---------------------------------------------------------------

\section{Results and discussion}

\subsection{Crystal structure}

\begin{figure}[htb]
\centering
\includegraphics[width=0.5\textwidth]{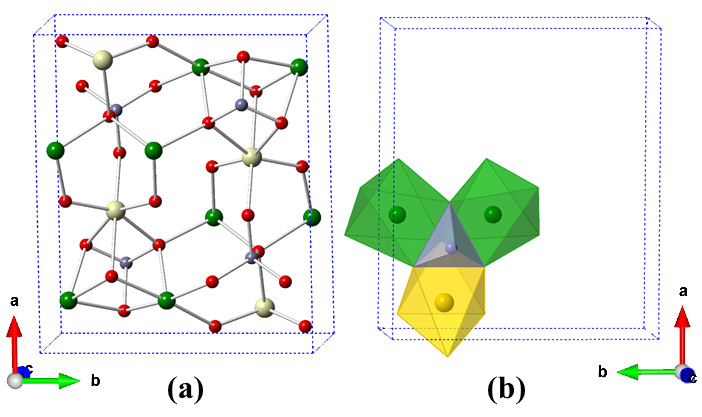}
\caption{(Color online) (a) A layer of edge-shared NiO$_6$ octahedra (green) shown in $c$-projection containing armchair chains along the $b$-direction, where every BO$_4$ tetrahedron (blue) in the neighboring layer bridges a pair of NiO$_6$ and one NbO$_6$ octahedra (yellow) through face-sharing, as shown in (b).} 
\label{fig:Figure2}
\end{figure}

The powder XRD pattern of the polycrystalline sample obtained from the pulverized 
as-grown Ni$_{2}$NbBO$_{6}$ single crystal sample is shown in Fig.~\ref{fig:Figure1}(a). 
All diffraction peaks can be indexed to the orthorhombic structure with space group 
Pnma, without any observable trace of impurity phase. The structural parameters 
were refined by the Rietveld technique with good quality refinement parameters (R$_{wp}$~=~1.67 \% and R$_{p}$~=~1.04 \%). 
The obtained values of the lattice parameters are \textit{a}~=~10.0690(1)~\AA, \textit{b}~=~8.6266(2)~\AA, and \textit{c}~=~4.4932(3)~\AA, which are in good 
agreement with  previously reported values~\cite{Ansell}. Fig.~\ref{fig:Figure1}(b) 
illustrates the single crystal XRD pattern with peaks indexed for the preferred orientation perpendicular to the (h00) planes. This compound could also be viewed as layers containing armchair chains of edge-shared NiO$_{6}$ octahedra, where each pair of NiO$_6$ along the $b$-direction are edge-shared with both NbO$_{6}$ octahedra and BO$_4$ tetrahedra, as illustrated in Fig.~\ref{fig:Figure2}. 

%---------------------------------------------------------------
%---------------------------------------------------------------

\subsection{Magnetic susceptibility}

%---------------------------------------------------------------
\begin{figure}[htb]
\centering
\includegraphics[width=0.45\textwidth]{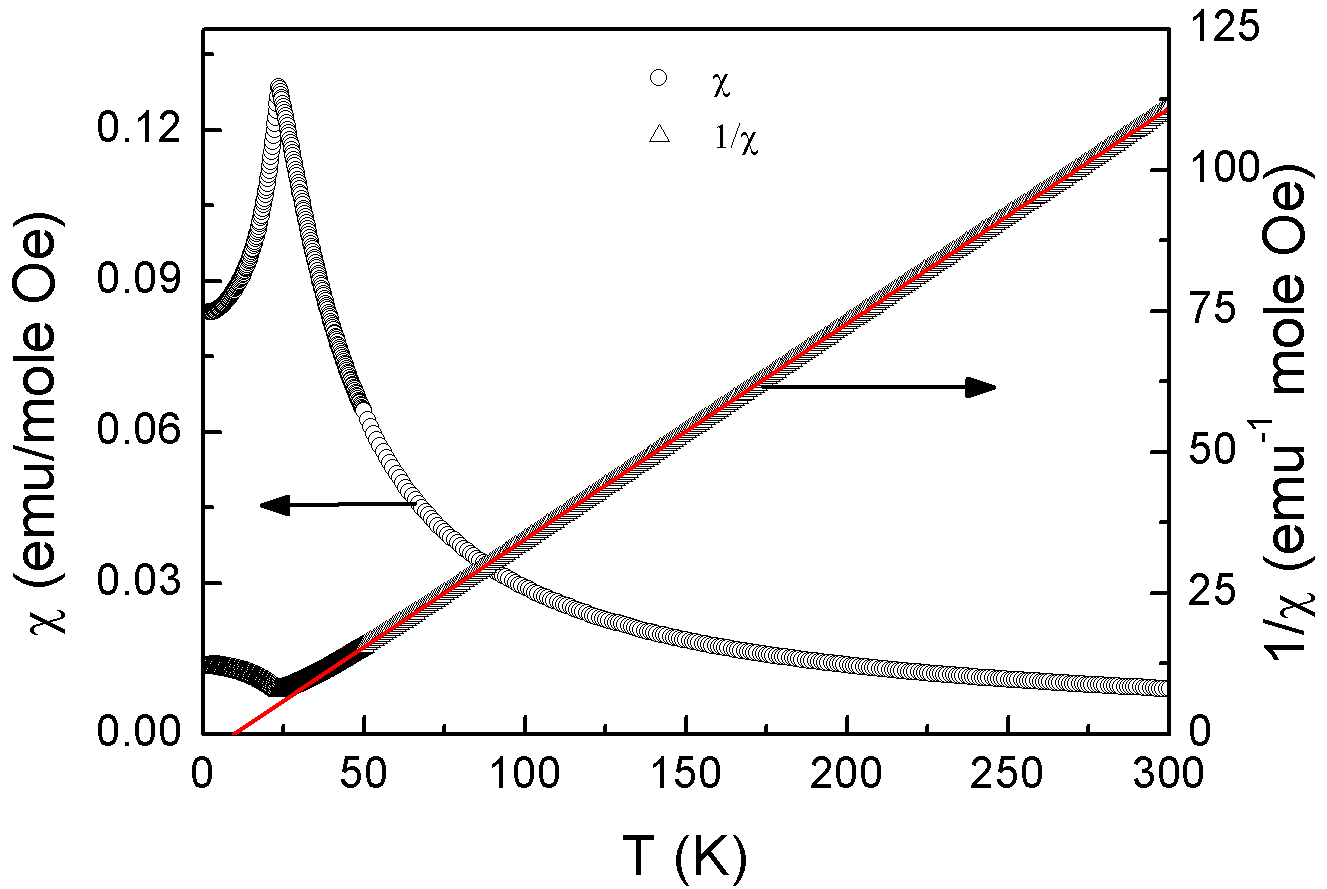}
\caption{(Color online) The temperature dependence of magnetic susceptibility 
$\chi$({\it T}) and the corresponding inverse susceptibility $\chi^{-1}({\it T})$ 
measured at an applied magnetic field of 10~kOe in the temperature range $2-300$~K for the 
pulverized as-grown Ni$_{2}$NbBO$_{6}$ crystals.}
\label{fig:Figure3}
\end{figure}
%---------------------------------------------------------------

Fig.~\ref{fig:Figure3} shows the temperature dependence of magnetic susceptibility 
$\chi$({\it T}) and the corresponding inverse susceptibility $\chi^{-1}({\it T})$ 
measured at an applied magnetic field of 10~kOe in the temperature range 2-300~K for 
pulverized powder of the as-grown Ni$_{2}$NbBO$_{6}$ crystal. The $\chi({\it T})$ curve 
shows a Curie-Weiss like behavior at high temperature and a sharp peak is observed at 24~K, 
indicating the onset of a antiferromagnetic ordering. The ordering temperature 
${\it T}_\mathrm{N}$~=~23.5~K is defined by the sharp peak through 
d($\chi${\it T})/d{\it T}. At ${\it T} > 50$~K, the $\chi({\it T})$ data 
can be fitted with the Curie-Weiss law ($\chi({\it T})$~=~{\it C}/({\it T}-$\theta$)) satisfactorily using the Curie constant {\it C}~=~1.31 and the Curie-Weiss temperature ($\theta$~=~9.5~K), as shown by the red solid line in Fig.~\ref{fig:Figure3}. 
The effective moment of $\mu_{eff}$ = 3.23~$\mu_{B}$ per Ni$^{2+}$ extracted from Curie constant is higher 
than the expected spin-only value of $\mu_{calc}$~=~2.83~$\mu_{B}$ for \textit{S}~=~1, which suggests the existence of a partially unquenched orbital contribution. The fitted value of $\theta$~=~9.5~K suggests the existence of an average ferromagnetic (FM) coupling among spins at high temperature but AFM ordering occurs at T$_N$$\sim$23.5~K, which suggests that the magnetic interactions must consider couplings beyond nearest neighbor spins and have different signs containing both ferromagnetic (FM) and AFM couplings, as verified later by our {\it ab initio} 
studies in the following. The $\chi$(\textit{T}) data above 200~K were also fitted by high temperature series (HTS) expansion up to 8$^{th}$ order.\cite{Rushbooke} The fitting parameters are found to be \textit{g}~=~2.02 and the exchange interaction (J/k$_B$)~=~-6.6 K. 

%---------------------------------------------------------------
\begin{figure}[htb]
\centering
\includegraphics[width=0.45\textwidth]{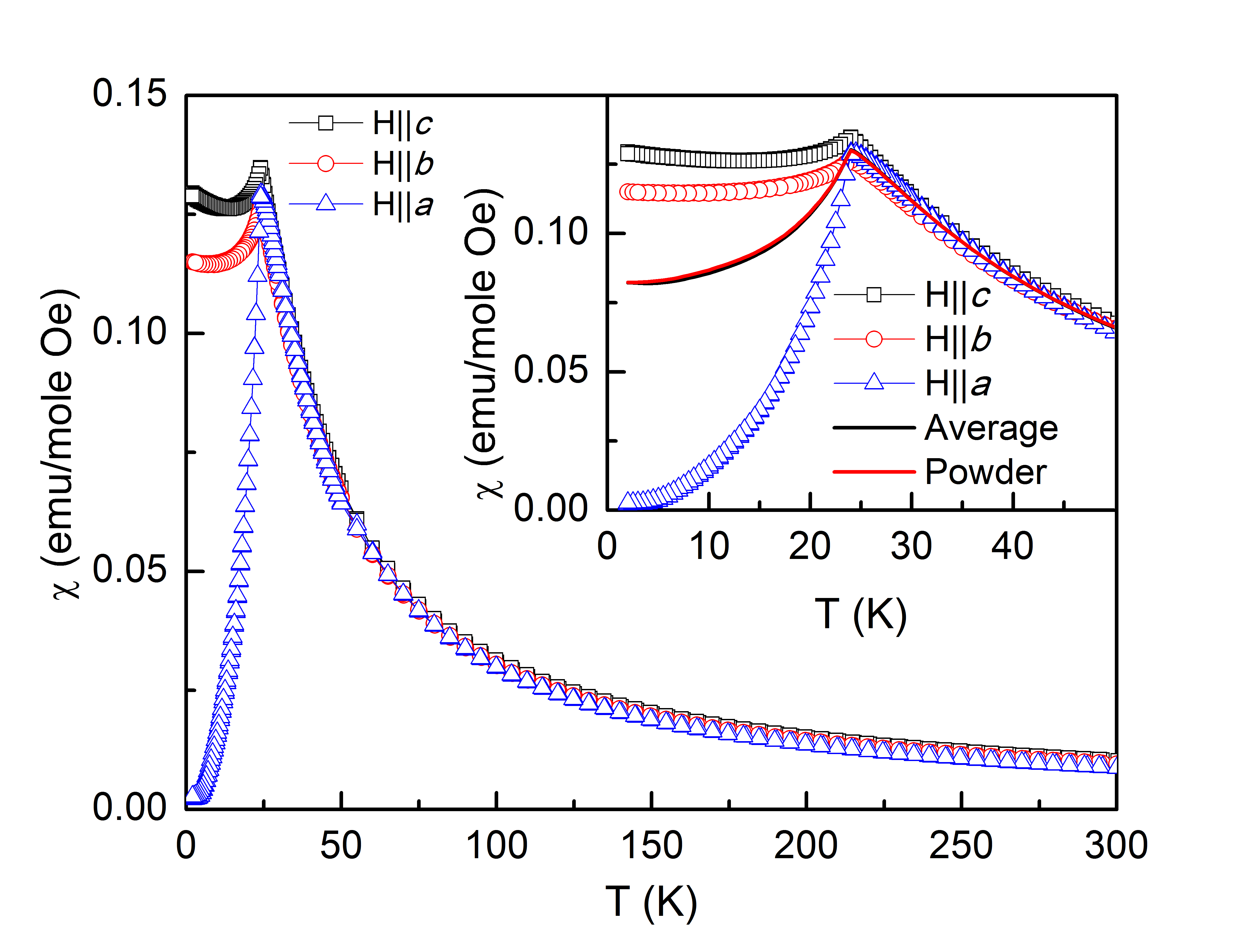}
\caption{(Color online) The magnetic susceptibilities $\chi$({\it T}) for 
Ni$_{2}$NbBO$_{6}$ single crystal measured in an applied magnetic field 
of 10~kOe parallel to all three crystallographic axis. The average of single crystal susceptibility ($\chi$({\it T})~=~($\chi_{a}$+$\chi_{b}$+$\chi_{c}$)/3) agrees perfectly with those measured using powder sample directly.}
\label{fig:Figure4}
\end{figure}
%---------------------------------------------------------------

Anisotropic magnetic susceptibilities $\chi$({\it T}) for Ni$_{2}$NbBO$_{6}$ single crystals were 
measured in an applied magnetic field of 10~kOe parallel to all three crystallographic 
axes, as shown in Fig.~\ref{fig:Figure4}. There is no deviation between 
$\chi_{ZFC}$({\it T}) and $\chi_{FC}$({\it T}) throughout the measured temperature range. 
Below $T_\mathrm{N}$, the anisotropy becomes significantly enhanced as shown in the 
inset of Fig.~\ref{fig:Figure4}, which indicates that the spins are aligned along the {\it a}-axis for in the 3D AFM long range ordering. 
%---------------------------------------------------------------
\begin{figure}[t]
\centering
\includegraphics[width=0.45\textwidth]{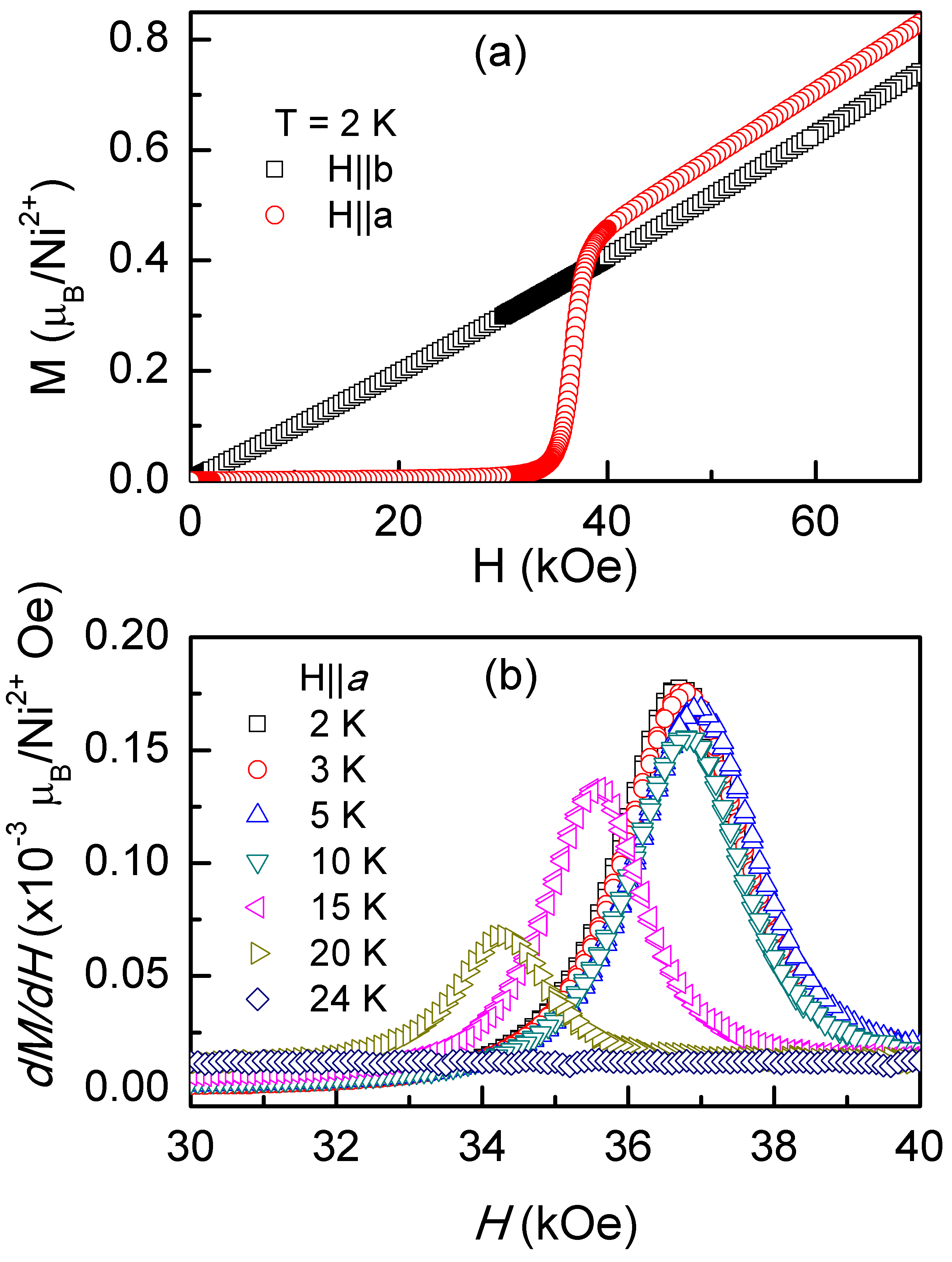}
\caption{(Color online) (a) Isothermal magnetization {\it M(H)} curves 
of Ni$_{2}$NbBO$_{6}$ for the magnetic field parallel to different 
crystallographic axis (b) d{\it M}/d{\it H} curves of 
Ni$_{2}$NbBO$_{6}$ for the magnetic field parallel to {\it a}-axis at 
some selected temperatures.}
\label{fig:Figure5}
\end{figure}
%---------------------------------------------------------------
%---------------------------------------------------------------
\begin{figure}[t]
\centering
\includegraphics[width=0.45\textwidth]{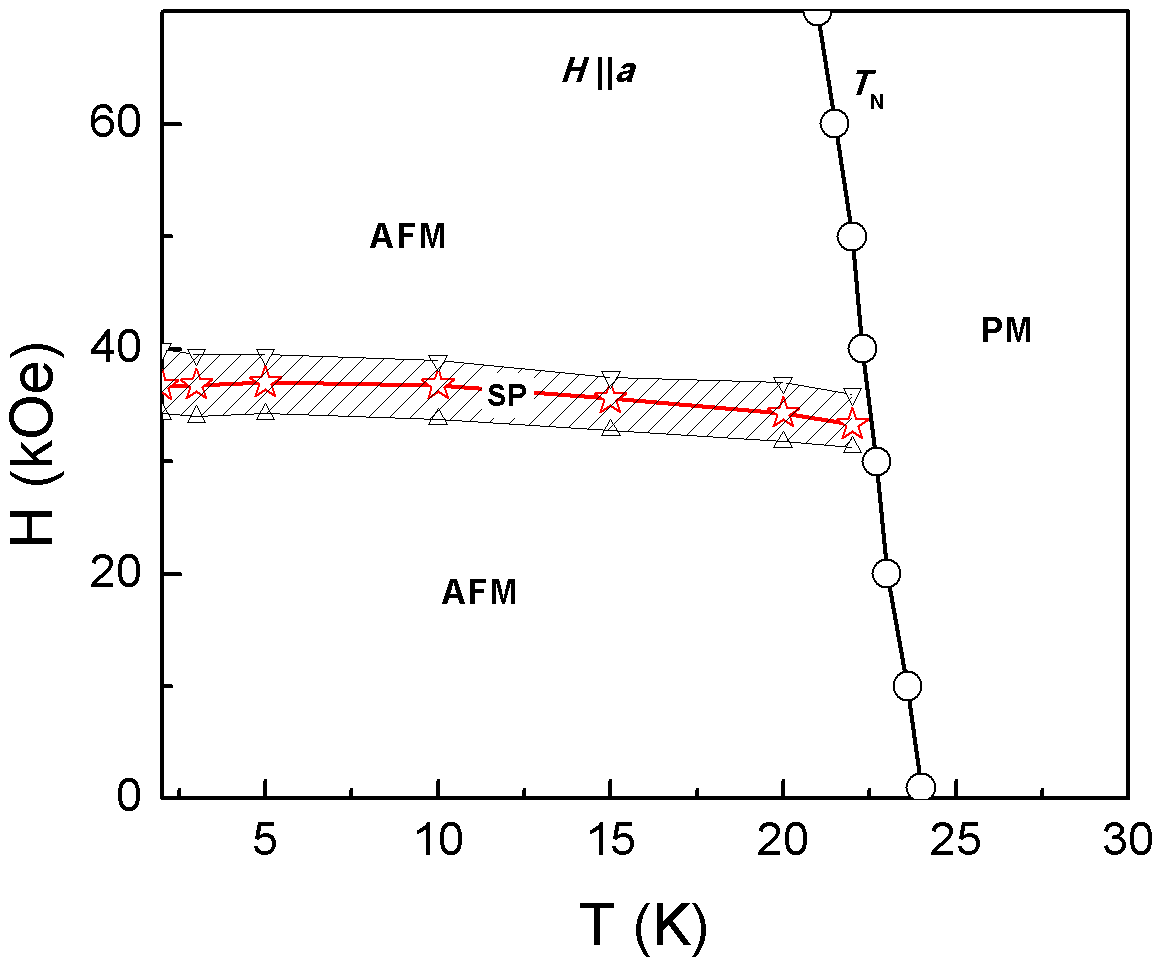}
\caption{(Color online) {\it H} - {\it T} phase diagram of 
Ni$_{2}$NbBO$_{6}$ for magnetic field parallel to {\it a}-axis 
obtained from magnetization measurements. The shaded area is related 
to the spin-flop (SP) transition defined by the start and end points of the 
transition from {\it M}({\it T},{\it H}). Open circles represent measured ${\it T}_\mathrm{N}$ and solid line is guide 
to the eye.}
\label{fig:Figure6}
\end{figure}
%---------------------------------------------------------------

To understand the AFM behavior of Ni$_{2}$NbBO$_{6}$ better, we 
measured magnetization as a function of magnetic field $H$ along the 
two crystal orientations above and below ${\it T}_\mathrm{N}$, as shown in Fig.~\ref{fig:Figure5}(a), where no field or temperature hysteresis were observed.  For magnetic field parallel to the {\it a}-axis, the magnetization reveals jump at a critical field {\it H}$_{\rm C}$ near {\it H}$_\mathrm{C}$~=~36.7~kOe below T$_N$ at 2~K, which is attributed to a spin-flop transition when the spin susceptibility changes abruptly to a higher level, {\it i.e.}, the ordered spins originally aligned along the $a$-direction flop to the direction perpendicular to the external field. As expected for the ordered spins aligned along the $a$-direction, we do not observe spin-flop transition manifested in {\it M(H)} for 
{\it H} is applied parallel to {\it b}- or {\it c}- axis.  We may summarize the {\it H-T} phase diagram for Ni$_{2}$NbBO$_{6}$ based on the magnetic field and temperature dependence of {\it M}({\it T,H}) with field applied 
parallel to {\it a}-axis, as shown in Fig.~\ref{fig:Figure6}. A small field dependence of ${\it T}_\mathrm{N}$ is also shown, where the boundary of spin-flop transition is indicated according to the onsets of d{\it M}/d{\it H} peaks shown in Fig.~\ref{fig:Figure5}(b). 

\subsection{Theoretical calculations}

\begin{figure}[t]
\centerline{
\includegraphics[width=8.50cm,angle=0]{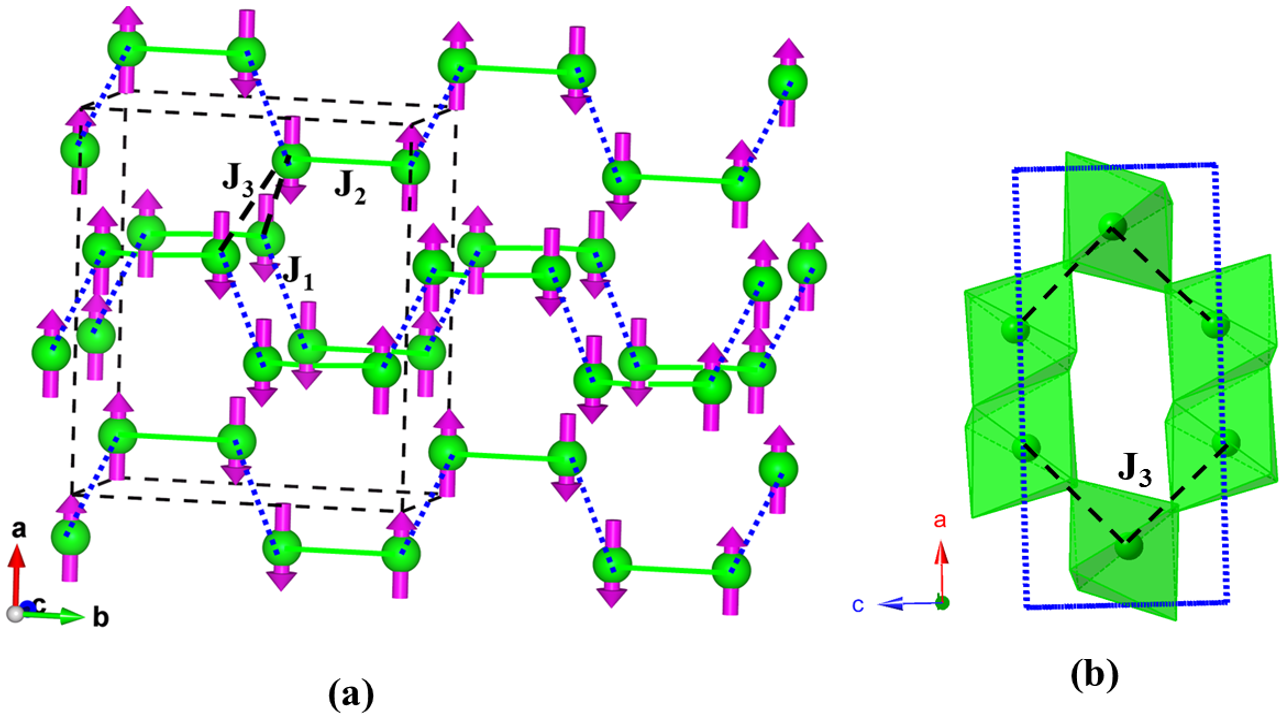}
}
\caption{(Color online) A schematic representation of the spin 
arrangement for configuration A.}
\label{fig:Figure7}
\end{figure}

Within first-principle density functional theory, 
we first calculated the total energy ($E_{FM}$) for the ferromagnetic state. 
The total energy per formula unit (f.u.) is -72.7578~eV. In order to find out the magnetic ground state of the 
system, we have considered various magnetic configurations
possible within the unit cell. Three configurations corresponding to the possible magnetic 
ground states have been used to estimate the exchange interactions, and three coupling 
constants ($J_i$) are considered based on the three shortest Ni-Ni distances, 
as shown in Fig.~\ref{fig:Figure7}. Here $J_1$, $J_2$, and $J_3$ represent the exchange
couplings between two neighboring Ni atoms corresponding to the Ni-Ni distances 
of 2.987~\AA, 3.099~\AA, and 3.436~\AA, respectively, as shown in Fig.~\ref{fig:Figure7}. 
Both ferromagnetic as well as antiferromagnetic alignments of Ni spins are considered and labelled as configurations A, B, and C. The calculated total energies of all these configurations are summarized in Table~\ref{tab:afm}~. 

We find that the configuration A (Fig.~\ref{fig:Figure7}) 
has the lowest energy, therefore configuration A is the expected magnetic ground state of the system. In this configuration, 
all the NN Ni ions within the armchair chain along the {\it b}-direction are antiferromagnetically coupled as a \textit{S}~=~1 dimer, and 
the other two NNN Ni ions, \textit{i.e.}, the inter-dimer within the armchair chain and the inter-chain couplings are ferromagnetically coupled. 

\begin{figure}[t]
\centerline{
\includegraphics[width=7.50cm,angle=0]{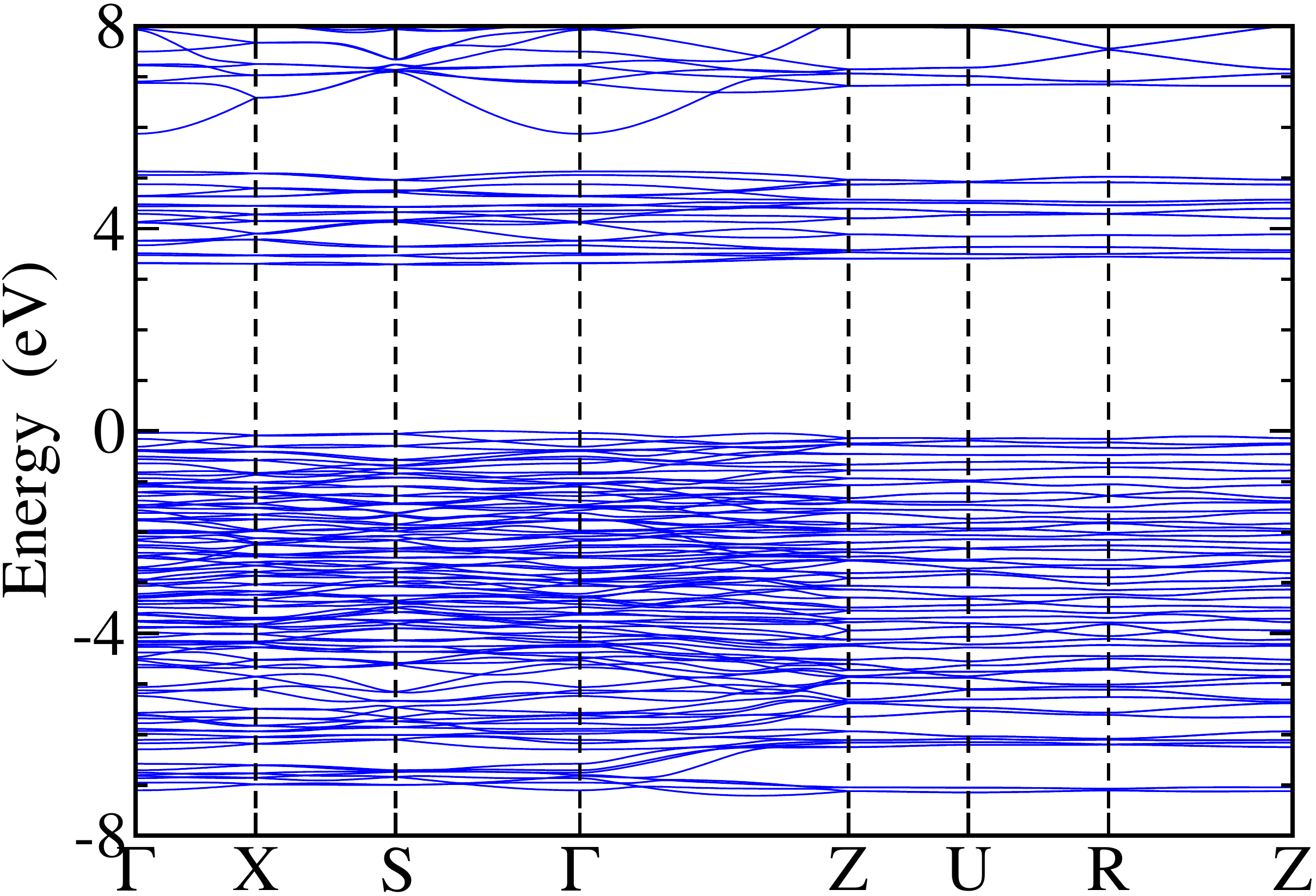}}
\centerline{
\includegraphics[width=7.50cm,angle=0]{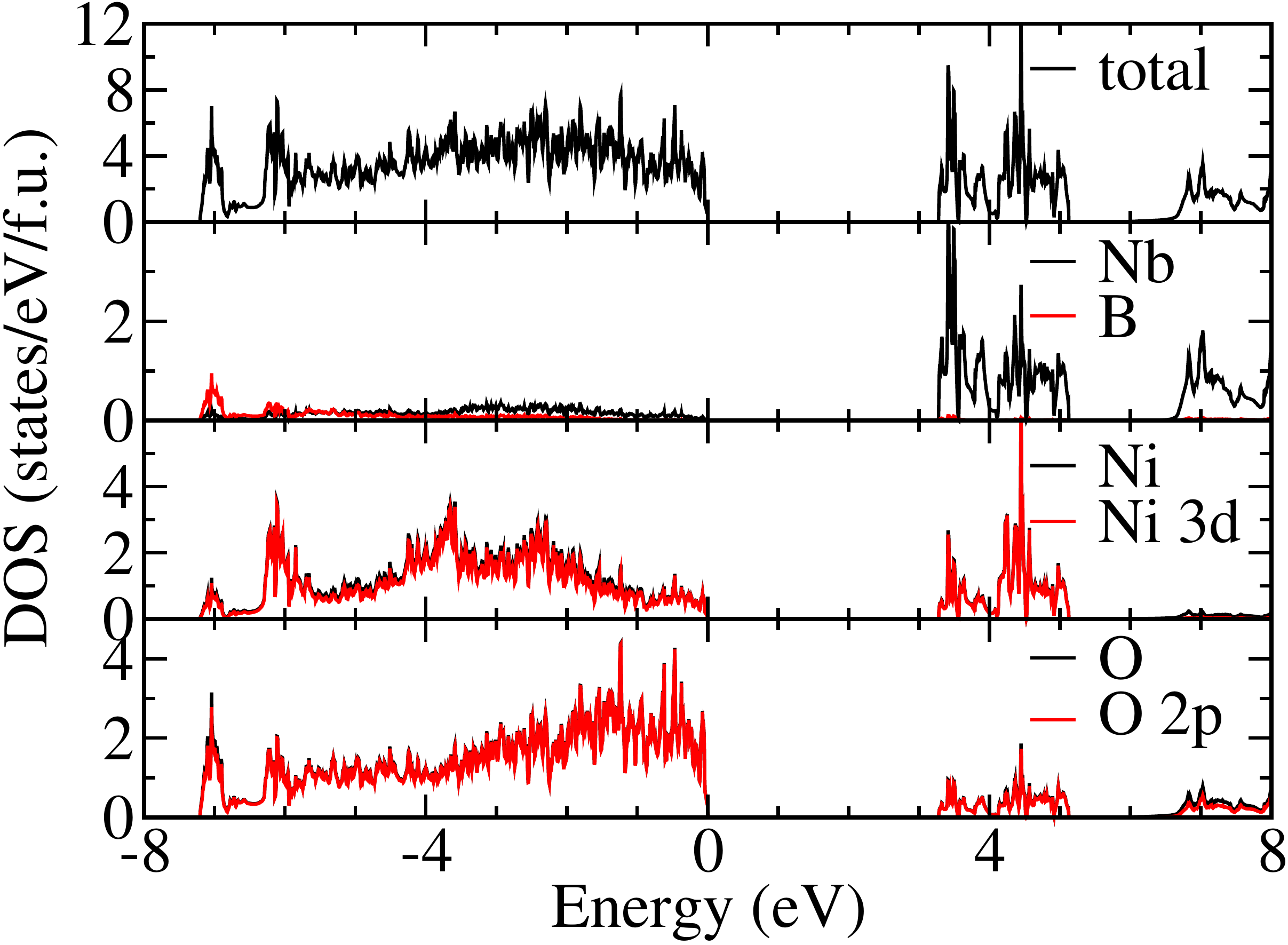}}
\caption{(Color online) Band structure (top panel) and density 
of states (bottom panel) of configuration A. Top of the valence 
band has been set to zero.}
\label{fig:Figure8}
\end{figure}

The calculated magnetic moment of Ni is $\sim$1.78~$\mu_B$, which is slightly smaller than the expected value of 2~$\mu_B$ for Ni$^{2+}$, suggesting that some of the magnetic moment lies outside the nickel atomic sphere used. Fig.~\ref{fig:Figure8} shows the band structure (top panel) and density of states (bottom panel) of configuration A. We find that the system has a large gap $\gtrsim$3 eV, which indicates that the system is insulating and consistent with the experimental observation. It is quite clear from the site-resolved density of states 
that the valence band is mainly composed of nickel 3{\it d} and oxygen 2{\it p} states. Thus the magnetic structure should be primarily decided by the spin-exchange coupling via Ni-O-Ni. 

%-----------------------------------------------------------------------------
%------------------------------------------------------------------------------
%------------------------------------------------------------------------------
\begin{table}[b]
\caption{Calculated total energy $\Delta E$ (relative to the total energy of FM state $E_{FM} = 72.7578$ eV/f.u.), 
total magnetic moment $m_s^{tot}$, atomic moment of Ni $m_s^{Ni}$ and band gap $E_g$}
\centering
\begin{tabular}{c rrrr}
\hline\hline
Config. & $\Delta E$  & $m_s^{tot}$ & $m_s^{Ni}$&$E_g$\\
        & (meV/f.u.) & ($\mu_B$/f.u.)& ($\mu_B$/atom)&(eV)\\[0.5ex]
\hline
FM      &  0.0     & 4.0 &  1.78 & 2.9\\
A       & -4.87    & 0.0 &  1.78 & 3.3\\
B       & -2.0     & 0.0 &  1.78 & 3.3 \\
C       & -2.16    & 0.0 &  1.78 & 3.3 \\[1ex]
\hline
\end{tabular}
\label{tab:afm}
\end{table}
%------------------------------------------------------------------------------

%-----------------------------------------------------------------------------
%-----------------------------------------------------------------------------

Next we evaluated the intra-chain and inter-chain 
magnetic exchange couplings among Ni spins. The exchange interaction 
between all the nearest-neighbor Ni ions along the 
{\it b}-direction in the $ab$-plane is denoted as $J_2$, and the other 
two couplings are $J_1$ (intra-chain) and $J_3$ (inter-chain), as illustrated in 
Fig.~\ref{fig:Figure7}.  In order to evaluate the exchange couplings, we have 
considered the obtained total energy of the unit cell 
of Ni$_2$NbBO$_6$ as the sum of the nearest-neighbor
spin-spin interactions in terms of the spin Heisenberg model
$H=E_0+\sum_{\langle ij\rangle} J_{ij} {\bf \sigma_i\cdot \sigma_j}$. 
Here $J_{ij}$ is the exchange interaction parameter between 
the nearest-neighbor Ni site $i$ and site $j$, and ${\bf \sigma_i}$ 
(${\bf \sigma_j}$) is the unit vector representing the direction 
of the local magnetic moment at site $i~(j)$.
The total energy per unit cell for all considered magnetic 
configurations are given by $E_{FM}=E_0+4(J_3+J_1)+4J_2$, 
$E_A=E_0+4(J_3+J_1)-4J_2$, $E_B=E_0+4(J_3-J_1)-4J_2$, and $E_C=E_0-4J_2$.
Solving the above mentioned equations we get the values of all 
exchange interactions listed in Table~\ref{tab:exint}~, where $J>0$ for AFM interaction and $J<0$ for FM interaction, and the constant $E_0$ contains all spin-independent interactions. 
Although the Ni-Ni distances for all these FM and AFM couplings are very close, the 
strength of the AFM coupling is nearly twice of the FM coupling, which could be due to the strong bonding of boron tetrahedra that bridge the NiO$_6$ pairs within each armchair chain. 
The AFM superexchange coupling has the largest magnitude of 2.43~meV, which roughly corresponds to ${\it T}_\mathrm{N}$$\sim$28~K and in good agreement with the experimental observation of ${\it T}_\mathrm{N}$$\sim$23.5~K.

%------------------------------------------------------------------------------
\begin{table}[t]
\caption{Calculated exchange interaction parameters (J$_{i}$) and the corresponding nearest-neighbor Ni-Ni distances.}
\centering
\begin{tabular}{c c c c}
\hline\hline
    & $J_1$ & $J_2$ & $J_3$  \\[0.5ex]
\hline
 J$_{i}$(meV) & -1.43 & 2.43 & -1.27  \\[1ex]
 Ni-Ni(\AA) & 2.987 & 3.099 & 3.436 \\[1ex]
\hline
\end{tabular}
\label{tab:exint}
\end{table}
%------------------------------------------------------------------------------

\section{Summary}

The magnetic properties of Ni$_{2}$NbBO$_{6}$ containing armchair chains have been studied in detail through {\it M(H,T)} measurement of the  
single crystal sample and compare with models predicted by the \textit{ab initio} calculations. A long range AFM spin ordering observed below ${\it T}_\mathrm{N}$$\sim$23.5~K. The spin-flop transition of critical field 36.7\,kOe at 2~K is found along {\it a}-axis and the \textit{H}-\textit{T} phase diagram is constructed accordingly. Within first-principle density functional theory, we have calculated the electronic and magnetic structures with exchange interactions that agree satisfactorily with the experimental results. We have established that Ni$_{2}$NbBO$_{6}$ consists of unusual armchair chains which are formed with \textit{S}~=~1 dimers with ferromagnetic intra- and inter-chain couplings.
 
\section{Acknowledgement}
FCC acknowledges the support provided by MOST-Taiwan under project number MOST 102-2119-M-002-004.
GYG acknowledges the financial support for this work from the Academia Sinica Thematic Research Program and 
the Ministry of Science and Technology of Taiwan.

\end{document}